# CONDENSATION OF SiC STARDUST IN CO NOVA OUTBURSTS

**Maitrayee Bose[1,*] & Sumner Starrfield[1]**

[1]School of Earth and Space Exploration, Arizona State University, Tempe AZ 85287-1404

*Center for Isotope Analysis (CIA), Arizona State University

Corresponding author: maitrayee.bose@asu.edu

## ABSTRACT

This study on presolar grains compares high-precision isotopic compositions of individual SiC grains with low $^{12}C/^{13}C$ ratios, low $^{14}N/^{15}N$ ratios, large $^{30}Si$ excesses and high $^{26}Al/^{27}Al$ ratios, available in the presolar grain database, to new CO nova models with white dwarf (WD) masses from 0.6–1.35$M_\odot$. The models were designed to match the Large Binocular Telescope (LBT) high dispersion spectra acquired for nova V5668 Sgr (Starrfield et al. 2018). These CO nova models provide elemental abundances up to calcium and include mixing of WD material into the accreted material in a binary star system under several scenarios, including one where mixing occurs only after temperatures $>7\times10^7$K are achieved during a thermonuclear runaway (TNR). The 0.8–1.35$M_\odot$ simulations where 25% of the WD core matter mixes with 75% of the accreted material (assumed solar) from its binary companion after the TNR has begun, provide the best fits to the measured isotopic data in four presolar grains. One grain matches the 50% accreted 50% solar 1.35$M_\odot$ simulation. For these five presolar grains, less than 25% of solar system material is required to be mixed with the CO nova ejecta to account for the grains' compositions. Thus, our study reports evidence of *pure* CO nova ejecta material in meteorites. Finally, we speculate that SiC grains can form in the winds of cool and dense CO novae, where the criterion C>O may not be locally imposed, and thus nova winds can be chemically inhomogeneous.

## 1. INTRODUCTION

Presolar grains are minute specks of rare dust grains in pristine meteorites. Tens of thousands of presolar SiC grains have been identified to date by their exotic isotopic compositions in major and trace elements (Zinner et al. 2014; Nittler and Ciesla 2016). These compositions are diagnostic of the gaseous environments in several stellar sites, where the dust grains condensed, and complement the radio and infrared observations of these stellar sites.

Presolar SiC grains are the most widely studied because they can be chemically extracted from meteorites and exhibit high concentrations of trace and rare earth elements. The SiC in both the large (1.8-3.7 μm) and small (0.5-0.65 μm) size fractions have been extensively investigated (e.g., Amari et al. 1994; Hoppe et al. 2010). Therefore, SiC grains provide unique opportunities to understand stellar evolution and nucleosynthesis, which has improved our understanding of processes in circumstellar envelopes and explosive environments, since their isolation in 1980s (Bernatowicz et al 1987; Anders and Zinner 1993; Amari et al. 1994). The C, N and Si isotopic compositions of the mainstream SiC grains and SiC X grains strongly affirm their origins in low-mass, solar metallicity, asymptotic giant branch stars, and core-collapse supernovae, respectively (e.g., Hoppe and Ott 1997; **Nittler et al. 1996**; Hoppe et al. 1996).

The stellar source has been under intense debate for some SiC grain types, however; for example SiC grains with low $^{12}C/^{13}C$ ratios <10, low $^{14}N/^{15}N$ isotope ratios in the range ~5-20, large $^{30}Si$ excesses and high $^{26}Al/^{27}Al$ ratios. The grains with these unique isotopic compositions were argued to have formed in ONe novae ("nova candidates", henceforth) (Amari et al. 2001) based on the model fits to several isotope systems (José et al. 1999; José and Hernanz 1998). The grain compositions require mixing of material synthesized in the nova outburst with isotopically close-to-solar system material. More specifically, the mixing of material between the white dwarf (WD) masses 1.15 and 1.25$M_\odot$ with >90% solar/terrestrial values were mandatory to account for the C, N, Si and Al isotopic ratios of the nova candidate grains (Amari et al. 2001). The implications from the Amari et al. (2001) study were 2-fold: first, the nova candidate grains were primarily from a binary system composed of the WD and a companion star on the main sequence. Second, the SiC grains probably formed from the outermost C-rich (C>O) ejecta blanket, very soon after the star exploded as a nova.

All these inferences about nova candidate grains are in stark contrast with what is known from observations of novae. First, CO novae are more abundant (70%–80%) than ONe novae and second, they are hypothesized to produce dust grains more efficiently (e.g., Gehrz et al. 1986; Mason et al. 1996).

Nittler and Hoppe (2005) disputed the origins of the nova candidate grains. They argued for supernova origins of some grains identified by Amari et al. (2001) that exhibited $^{47,49}$Ti anomalies and $^{44}$Ca excesses. For example, one grain 334-2 (Nittler and Hoppe 2005) that exhibits low $^{12}$C/$^{13}$C and $^{14}$N/$^{15}$N ratios has Si and Ca isotopic compositions akin to SiC X grains. $^{44}$Ca enrichment in this grain points toward the initial presence of $^{44}$Ti, only synthesized in supernovae. All these factors combined indicate that this particular grain 334-2 originally attributed to have a nova origin, should be reclassified as a supernova grain. Another grain 151-4 argued to have an origin in a ~1.2$M_\odot$ ONe nova exhibits $^{47}$Ti excesses. Although the origins of $^{47}$Ti is uncertain, $^{47}$Ti anomalies can originate in both Type II supernovae (Amari and Zinner 1997) and Type Ia supernovae (Woosley and Weaver 1994). Nittler and Hoppe (2005) argued that the $^{47}$Ti excesses or, in general, Ti isotopic anomalies can be produced only at high temperatures that may not be attained in nova explosions. The study inferred that grain 151-4 with $^{47}$Ti excesses formed in an astrophysical setting, other than novae. This study clearly indicates the need for measuring multiple isotopic systems in the same grain, and a comparison between models and laboratory measurements can yield improved results when the number of isotopic systems measured in the same grain are large.

New nova simulations using a more precise thermonuclear rate of the $^{33}$S(p,γ)$^{34}$S constrained the $^{32}$S/$^{33}$S ratios in nova models to 110-130, whereas recent type II supernova models predict $^{32}$S/$^{33}$S ratios of 130–200 so sulfur isotopes were claimed as the means to distinguish between grains of nova and supernova origin (Parikh et al. 2014). This hypothesis led to measurement of sulfur isotopes in **seven** nova candidate grains, none of which exhibit $^{33}$S/$^{32}$S ratios in the range speculated by Parikh et al. (2014) (**$^{32}$S/$^{33}$S** = 110–130). Three grains show depletions of $^{33}$S, while one shows δ$^{33}$S/$^{32}$S = 48±334 (Liu et al. 2016; Delta notation is defined as δ$^{33}$S/$^{32}$S = ($^{33}$S/$^{32}$S$_{sample}$/$^{33}$S/$^{32}$S$_{Earth}$-1)×1000‰). Lower than solar $^{34}$S/$^{32}$S ratios in two of them (δ$^{33}$S/$^{32}$S=-542±175; -394±106) was used as evidence for ruling out their origins in ONe novae (Liu et al. 2016). Liu et al. (2016) further discussed that several nova candidate grains could have their origins in Type II supernovae, and an unambiguous assignment may never be possible because proton capture in both these explosive H burning environments would produce p-rich radionuclides. More recently, Illiadis et al. (2018) used a different strategy to constrain the origin of nova grains by simulating nova grain compositions over a large parameter space, including reaction rates, peak temperature, density, and decay time. A comparison of these simulated compositions to the presolar grain measurements show that only a small subset (16%) of nova candidate grains exhibits isotopic signatures consistent with a nova origin. This strategy works better for grains that have more isotopic ratios measured, i.e., the larger the number of isotope ratios known, the higher the chance of getting an acceptable solution to their simulations.

In this paper, we use the isotopic abundances from recent CO nova simulations from Starrfield et al. (2018) to compare to the existing database of putative nova grains. Because these simulations attempt to match the peak luminosities and ejection velocities observed in nova explosions, they are more diagnostic of actual environments. The differences between the models described in Iliadis et al. (2018) and Starrfield et al. (2018) are discussed in Section 2. Some SiC nova candidate grain compositions can be well explained by these new CO nova simulations, and therefore CO novae could have contributed to the presolar grain inventory that was injected into the protosolar molecular cloud. Carbonaceous dust grains have been observed to form O-rich binary stars, although the premise C>O may not be satisfied. All types of dust (SiC, silicates, hydrocarbons) form at different times during a single nova outburst probably because CO formation does not go to completion and nova ejecta has large abundance gradients (Gehrz et al. 1992).

## 2. NOVA MODELS

The calculations reported in Starrfield et al. (2018) were done using NOVA. NOVA is a one-dimensional, fully implicit, Lagrangian hydrodynamic computer code described in Starrfield et al. (2009, and references therein). The simulations that produced the isotopic abundances used in this paper were done with 150 mass zones and convective mixing was done with the Arnett et al. (2010) algorithm. They used a mixing-length to scale height ratio of 4. In contrast to earlier work, they used the Starlib reaction rate library (Sallaska et al. 2013). The simulations were done for CO (carbon 50% by mass and oxygen 50% by mass) white dwarf masses of $0.6M_\odot$, $0.8M_\odot$, $1.0M_\odot$, $1.15M_\odot$, $1.25M_\odot$, and $1.35M_\odot$ with a mass accretion rate of $2 \times 10^{-10}$ $M_\odot$ yr$^{-1}$ and an initial luminosity of $4 \times 10^{-3}$ $L_\odot$. For all cases the tabulated abundances were obtained from that fraction of the envelope that, as a result of the TNR, had reached escape velocity and was optically thin. There are 3 sets of simulations reported in Starrfield et al. (2018). The initial white dwarf structure was the same for all the simulations only the composition of the accreting material was changed. The first set of simulations assumed that the accreting material had mixed with the white dwarf material from the beginning (MFB). This is the same assumption as in all the previous studies by Starrfield and collaborators. Two separate compositions were used (all abundances are mass fractions): either 25% core material and 75% Solar (Lodders 2003) matter (25-75) or 50% core material and 50% Solar (50-50). These simulations were followed through the peak of the TNR and for a sufficiently long time after to determine the amount of ejected material and its velocity. These simulations (MFB 25-75 and MFB 50-50 are plotted in the Figures 1 and 2) were in poor agreement with the carbon, nitrogen and silicon isotope data in SiC grains. In addition, these simulations don't eject sufficient material with significant velocities to agree with the observations of nova explosions (Bode and Evans 2008).

The second set of simulations involved accreting just Solar material and these were also followed through the explosion. The third set assumed that mixing of the core with envelope did not occur until the peak temperature in the TNR (initiated with the pure Solar mixture) had reached about $7 \times 10^7$K. At this time the composition of the accreted layers was instantaneously switched to either the 25-75 mixture or the 50-50 mixture. As reported in Starrfield et al. (2018), the simulations with the new mixture took only a few seconds to adjust and the resulting structure was followed through the peak of the TNR and the determination of the amount of ejected matter and velocities. The isotopic abundances in the ejected material were then tabulated, used for the studies in the current work, and reported in Starrfield et al. (2018). The initial $^{12}$C abundance (from Lodders and Palme 2009) makes a difference when the TNR is reached and the simulations with more $^{12}$C evolve faster with less accreted mass. The mixed core-accreted compositions in our simulations are composed of either 25% WD core and 75% solar material or 50% WD core and 50% accreted material and are referred to as MDTNR 25-75 and MDTNR 50-50, respectively. The temperatures reached in the deepest H-rich zones increases with the WD mass, and for the 1.35$M_\odot$ MDTNR 25-75 and MDTNR 50-50 models are $3.4\times10^8$K and $6\times10^8$K, respectively (Starrfield et al. 2018). Furthermore, the amount of $^{12}$C is less in the MDTNR 25-75 simulations, which results in longer decline to quiescence.

Iliadis et al (2018) used a totally different procedure and it is difficult to compare its results with our study. Instead of doing a full evolutionary simulation, they choose values for the peak temperature and density, and then assume an exponential decrease in those values as a function of time. The study assumed a white dwarf mass of 1.0$M_\odot$, luminosity of $10^{-2}$L$_\odot$, accreted solar-like material at a mass accretion rate of 2 x $10^{-10}$ $M_\odot$ yr$^{-1}$ and assumed MFB with a 50-50 mixture. Iliadis et al. (2018) did compare their results to one simulation done with SHIVA (Jose and Hernanz 1998) and achieved reasonable agreement (factor of 2) between their method and the SHIVA study (see Figure 2 of Iliadis et al. 2018). They then varied both the reaction rates and the values of temperature, density, and decay time using a Monte Carlo technique; compared their isotopic results to nova candidate grain compositions; and were able to find reasonable agreement with some grains (by mixing the ejecta "with more than 10 times the amount of unprocessed, solar-like matter before grain condensation" (Iliadis et al. 2018)).

The general trends observed in the Starrfield et al. (2018) CO nova simulations are as follows: The MDTNR simulations produce low $^{12}$C/$^{13}$C ratios (0.03–3.4) and variable $^{14}$N/$^{15}$N ratios (up to 243) that decrease with increasing WD mass for both mixtures. The MDTNR values show similar $^{14}$N/$^{15}$N values for >0.8$M_\odot$ WDs. For Si isotopes, with an increase in the WD mass, significant $^{30}$Si excesses are observed, with the $^{28}$Si/$^{30}$Si ratios <16 for 1.15–1.35$M_\odot$ WD models. The $^{28}$Si/$^{29}$Si ratios, however, show an enhanced ratio for the 1.15$M_\odot$ WD (50–53), followed by some enhancement in $^{29}$Si. In the case of sulfur isotopes, $^{33}$S enrichments are observed in >1.15$M_\odot$ producing the lowest $^{32}$S/$^{33}$S ratios ~15. The $^{32}$S/$^{34}$S ratios are overall constant with the lowest $^{32}$S/$^{34}$S ratio of ~10 but show a spike ($^{32}$S/$^{34}$S ~ 84–108) for the 1.25$M_\odot$ WD. The $^{26}$Al/$^{27}$Al ratios of the models show a large range (0.0054 – 0.55) with a limited range 0.11–0.27 in >1.15$M_\odot$ WDs. The $^{26}$Al/$^{27}$Al ratios are high for 1.15$M_\odot$ WD with ratios up to 0.55.

## 3. COMPARISONS WITH PRESOLAR GRAINS

Table 1 lists the carbon and nitrogen isotope ratios of thirty presolar SiC grains obtained from the literature (Hoppe et al. 1996; Gao and Nittler 1997; Amari et al. 2001; Nittler and Alexander 2003; Nittler and Hoppe 2005; Liu et al. 2016, 2017; Hoppe et al. 2018). Three of those grains don't have either carbon or nitrogen isotopic compositions and so Figure 1 shows only the compositions of twenty seven presolar SiC grains, including six SiC grains with the highest probability of being nova condensates (Iliadis et al. 2018) in orange. The theoretically predicted CO models are plotted in different symbols. Except for the 0.6$M_\odot$ MFB 50-50 simulation, all remaining CO nova simulations plot in the 3$^{rd}$ quadrant (lower left) of the carbon-nitrogen isotope space (Figure 1).

These new CO results are in stark contrast to the predictions from prior studies (e.g., José et al. 2004). CO results from José et al. (2004) spanned a large range in $^{14}N/^{15}N$ ratios from 5 to 50000. ONe compositions from José et al. (2004), on the other hand, plot approximately at the location of the CO model simulations described here.

The presolar SiC data plot in between the terrestrial ratios and compositions of the MDTNR CO nova simulations predominantly in the lower left in Figure 1. Mixing between the products of nucleosynthesis in the nova explosion and isotopically close-to-solar material is thus required to explain the grains compositions. Mixing lines for the MDTNR 25-75 and 50-50 models with 0.8–1.35$M_{\odot}$ WD masses are shown in Figure 1. We consider that a grain composition matches a simulation if the mixing lines fall within <2–4 times the errors on the measured compositions. Based on this assumption, the MDTNR 25-75 and 50-50 models with WD mass 1.00–1.35$M_{\odot}$ can quantitatively explain the carbon and nitrogen isotopic data for a majority (17 out of 27) of the SiC grains (without yellow outlines in Figure 1), including four grains with the highest probability of being nova grains (Iliadis et al. 2018). In several cases, simulations with different WD masses can explain a grains' carbon and nitrogen isotopic compositions simultaneously (Figure 1). Table 2 shows all the models that can reproduce the grain's compositions. The individual cells in Table 2 are either in bold or in italics to identify grains where 4–5 isotope ratios fit the simulations. The grains whose compositions can be explained by the mixing lines mostly require a larger proportion of the material from the nova. The proportion of nova ejecta varies from ~80% to >95% for a majority of the grains, in sharp contrast to previous work. Previous work required <5% of the material to be from the nova ejecta (e.g., Amari et al. 2001). The proportion of nova ejecta to be mixed with the solar system material is listed (in percentage) for the MDTNR 25-75 simulation with 1.15$M_{\odot}$ WD mass because it can explain a large number of nova candidate compositions (Figure 1). Several grains that lie on the 1.15$M_{\odot}$ mixing line require 90% material from the nova ejecta. The highly plausible grain M11-151-4 that plots on the MDTNR 25-75 1.15$M_{\odot}$ mixing line require ~95% of material from the nova ejecta. Finally, **eleven** SiC grains (black and orange symbols with yellow outlines in Figure 1) cannot be explained by any of the CO models. These include AF15bC-126-3, G240-1, G1697, Ag2, Ag2_6, G270_2, M2-A4-G672, M2-A5-G1211, M1-A8-G145, KJD-1-11-5 and KJD-3-23-5 (Table 2).

The consistency between the grain data and simulations qualitatively does not break down when we consider other isotopic systems. A comparison of the silicon isotopes of twenty nine SiC grains (Hoppe et al. 1996; Gao and Nittler 1997; Amari et al. 2001; Nittler and Alexander 2003; Nittler and Hoppe 2005; Liu et al. 2016, 2017; Hoppe et al. 2018) and nova models indicate that ten SiC grains with close to solar $^{28}Si/^{29}Si$ ratios and some $^{30}Si$ enrichments match the silicon isotope compositions of the low-mass (0.6–0.8$M_{\odot}$) MDTNR simulations (Figure 2, 2-inset, Table 2). More precisely, these SiC grains can be explained by the MDTNR 25-75 CO simulations with *no mixing* between the nova and solar system materials. Five SiC grains with correlated $^{29}Si$- and $^{30}Si$-enrichments (3rd quadrant of the Si isotope plot) fall on the MDTNR 50-50 model with the high WD mass of 1.35$M_{\odot}$. These grains require variable amounts of nova ejecta from ~50-80% (Figure 2). Eight SiC grains with $^{30}Si$ excesses and $^{29}Si$ depletion (2nd quadrant of the silicon isotope plot) can be very well explained by several simulations including MDTNR 25-75 models with WD masses 1.00$M_{\odot}$, 1.15$M_{\odot}$ and 1.25$M_{\odot}$. For example, the highly plausible SiC grain M11-151-4 composition requires mixing 75% of the nova ejecta with 25% of solar system material in the MDTNR 25-75 simulation with 1.15$M_{\odot}$ WD (Figure 2). Finally, the remaining four grains cannot be explained quantitatively by any of the simulations, including G270_2, M2-A4-G27, AF15bB-429-3, M2-A5-G1211 (black and orange symbols with yellow outlines in Figure 2).

Figure 3 shows the sulfur isotopic compositions of six SiC grains, two of which are included in the plausible nova candidate grains based on Iliadis et al. (2018). One grain SM1-A8-G410 is outside the range in Figure 3 with very large $^{33}S$ depletions ($\delta^{33}S$ = -833 ‰, Table 1). We observe that several MDTNR models with WD masses from 0.6–1.15$M_{\odot}$ have sulfur compositions exactly consistent with three SiC grains (i.e., *no mixing* required) as shown in Figure 3-inset. One grain Ag2_6 (Figure 3) with $^{34}S$ depletion and normal $^{32}S/^{33}S$ ratios can be explained by mixing of solar system material with <50% of nova ejecta in both MDTNR 25-75 and 50-50 models with 1.25$M_{\odot}$ WD, owing to the large uncertainties in the grain measurement. Two remaining grains, G270_2 (highly plausible nova candidate by Iliadis et al. 2018) and KJD-1-11-5 (Figure 3; Table 2) with some $^{33}S$ and $^{34}S$ depletions cannot be explained by any CO simulations.

Finally, the $^{26}Al/^{27}Al$ ratios of seventeen SiC grains span a large range from $1.2\times10^{-2}$ to $3.9\times10^{-1}$ (Figure 4). About 47% (eight grains) of these grains can be quantitatively described by the MDTNR simulations described here, including two highly plausible grains from Iliadis et al. (2018). The remaining grains don’t fit the simulations (Figure 4, Table 2). The grains whose compositions can be quantitatively explained by high-mass models, e.g., 1.35$M_{\odot}$ WD MDTNR model require ~85–95% nova material to be mixed with solar system material, which agrees well with observations in other isotopes.

Therefore, based on the simple comparison of the presolar grain data to the new models, we are able to identify several simulations that provide good fits. The good fits for each grain are listed in Table 2. Table 2 shows the grains in different shades of green with the lightest shade for grains where two isotopic ratios provide a good match to the simulations and the darkest shade for grains where models provide best fits to five isotopic ratios. This enables us to do a meaningful comparison of the grain data to the simulations on a grain-by-grain basis. The technique we used to identify the *true* nova grains based on such detailed comparisons is discussed below. The uncertainties in nucleosynthetic simulations can be large, and so we include the simulations that could give good fits given 2–4 times the error on the measurement. It is clear from Table 2 that several simulations are able to reproduce each grain composition. The other factor that we are aware of are the systematic errors associated with sample preparation and difficulties during isotopic measurements of contaminated grains. For example, sulfur isotope data of presolar SiC grains can be contaminated with terrestrial and meteoritic contributions. In addition, significant quantities of sulfur are not expected to condense into SiC. The observed $^{32}S$ excesses in SiC X grains of supernova origin are likely due to decay of radioactive $^{32}Si$ (Pignatari et al. 2013). This pure $^{32}S$ component is diluted by sulfur contamination that must have lowered the true anomalies. Inspite of these uncertainties, the sulfur isotopic compositions of three SiC grains match the compositions of the simulations well (Figure 3, Table 2).

## 4. DISCUSSION

***4.1. Nova grains in the presolar grain inventory.*** Because any meaningful comparison of stellar model predictions with presolar grains should be based on multi-element isotope data of individual grains, we list all simulations that fit carbon, nitrogen, silicon, sulfur, and aluminum isotope data simultaneously for all the grains in Table 2. Getting a quantitative solution considering carbon and nitrogen only was not difficult but getting a good fit to the silicon data for the same simulation did not occur in the majority of cases. Only 8 grain compositions can be reproduced by at least four isotopic ratios and they have been included in Table 3. Some of these grains require several different scenarios. Next, we did mixing calculations between the terrestrial ratios and nova compositions and calculated the contribution from the nova ejecta in each case. The contribution from the nova ejecta that will explain the SiC isotopic compositions is listed in Table 3. We consider each of these 8 grains in more detail below. We designate a grain to be a nova condensate, if the proportion of nova contribution from the mixing calculations is similar (within 25%) for any three isotope ratios. Note that the dispersion in the output of the model simulations are unknown at present but can be as large as 50% (e.g., by changing the WD mass), and so grains with proportions of nova contribution that differ by 50% are considered plausible nova grains (*maybe* category). This is defined as our criteria for the discussion below.

***4.1.1. M11-151-4 and M2-A1-G410.*** These two grains fit the MDTNR 25-75 simulations, with WD mass between 0.8–1.15$M_\odot$. Both require >88% matter from the nova to explain their carbon and nitrogen isotope compositions. M11-151-4 requires 75% of nova contribution to match silicon isotopes and 96% for $^{26}Al/^{27}Al$ isotope ratio, which certainly makes this grain a nova grain. In case of the grain M2-A1-G410, two simulations (Table 3) gave promising results. The MDTNR 25-75 1.00$M_\odot$ simulation required a 20% contribution from the nova to explain the grains' silicon isotopic composition, which was used to rule out this scenario. Alternatively, the simulation MDTNR 25-75 0.8$M_\odot$ matches the carbon, nitrogen, silicon and aluminum isotope ratios very well, with contributions from the nova greater than 90%. Neither of these grains have sulfur isotope data but considering that **five** isotope ratios are well matched for the MDTNR 25-75 0.8$M_\odot$ simulation, our criteria suggests that these are produced in novae ejecta.

Grain M11-151-4 shows $^{47}Ti$ excesses, which were interpreted by Nittler and Hoppe (2005) to be a supernova signature because of the inability of nova to reach high enough temperatures. However, our 1.35$M_\odot$ simulations are able to achieve a sufficiently high temperature that makes production of $^{47}Ti$ possible.

***4.1.2. G1342.*** The simulation MDTNR 50-50 with a high mass (1.35$M_\odot$) WD explains the carbon and nitrogen isotope compositions of this grain and requires 98% contribution from the nova. This grain has no sulfur data and we were unable to get a good fit for aluminum from this simulation. The fit for silicon isotopes works but the percentage contribution from the nova is about 55%. Although the uncertainties in the model parameters can be large (~50%), we consider this grain in the 'maybe' category, keeping our criteria in mind.

***4.1.3. GAB.*** This grain matches both MDTNR 25-75 and MDTNR 50-50 simulations with a WD mass of 1.00$M_{\odot}$. For both simulations, the nova contribution that explains the carbon and nitrogen isotopic compositions is ~98% and sulfur isotope ratios is 100%. Although, no good fits for silicon or aluminum isotope ratios could be attained in these simulations, the fact that almost identical amounts of material is required from the nova simulation makes it a nova grain. Note that this is the only grain in our list (Table 3) where two simulations are able to reproduce the grains' composition well.

***4.1.4. G283.*** The carbon and nitrogen isotopic compositions of grain G283 can be well explained by four different simulations, MDTNR 25-75 1.00$M_{\odot}$, MDTNR 25-75 1.25$M_{\odot}$, MDTNR 25-75 1.35$M_{\odot}$, and MDTNR 50-50 1.25$M_{\odot}$. Only the MDTNR 25-75 1.35$M_{\odot}$ simulation provides a good fit to the carbon, nitrogen, and aluminum isotope ratios. The remaining simulations do not provide good fits to the aluminum isotope ratios and the contribution for silicon isotopes is <20% in all cases. Silicon is a major element in the grain and the inability of the simulations to provide a good fit is not understood. But based on our criteria, we consider G283 a nova grain, considering our criterion.

***4.1.5. G1748, M2-A1-G114, and M2-A5-G269.*** For these three grains, the contribution of >80% nova matter is necessary to explain the carbon and nitrogen isotopic compositions. However, for silicon, the same models only require ~10-20% of matter from the nova ejecta. Sulfur and aluminum isotopes have not been measured for M2-A1-G114 and M2-A5-G269. Furthermore, a good fit for aluminum isotope ratios in grain G1748 could not be achieved. Therefore, we do not consider these grains to be nova grains, and other stellar sites need to be considered.

Therefore, this detailed comparison between the nova models and grain compositions has allowed us to rule out several nova candidate grains and identify four grains, namely M11-151-4, M2-A1-G410, GAB, and G283 that are likely *true* nova condensates. SiC grain G1342 is currently placed in the 'maybe category', and uncertainties in models need to be investigated to ascertain its origins. Only grain M11-151-4 from Nittler and Hoppe (2005) is included in Iliadis et al. (2018) as a high probability CO nova grain. Four of these can be explained by 25-75 MDTNR models with WD masses of 0.8, 1.00, 1.15, and 1.35$M_\odot$. The MDTNR 50-50 simulation in the high WD mass range (1.35$M_\odot$) works well too.

***4.2. Carbon-rich grain formation in CO novae.*** Although earlier work (José et al. 2004; José and Hernanz 2007) had shown that CO novae should exhibit limited nuclear activity beyond the CNO region because of the moderate peak temperatures achieved during the explosion, and lack of seed nuclei of the heavier masses, the new models presented here contradicts those results. The primary differences between the two studies (Starrfield et al. 2018; Iliadis et al. 2018) was presented earlier. High $^{12}C/^{1}H$ ratios result in an increase in the rate of energy production during the CNO cycle, which in turn produces high temperatures in the nuclear burning region (Starrfield et al. 2018). Thus, good agreement between the new MDTNR CO nova simulations and a small selection (<20%) of the presolar SiC grain data are achieved and nova grains form a small, yet significant fraction of the presolar grain inventory.

Equilibrium condensation calculations require that the carbon-rich grains condense if C>O and oxide/silicate grains condense if C<O. Because both oxide and carbon-bearing dust grains have been reported through IR spectroscopy of nova outbursts (Gehrz et al. 1992, 1995; Mason et al 1996, 1997), it may imply that locally the C<O criterion *may not* be met, allowing for chemical heterogeneity in the nova ejecta. It was suggested that oxygen-rich supernova environments can form graphite stardust, depending strongly on the density of the CO-bearing gas and production of free carbon by thermal radiation and heating by radionuclides (Denealt et al. 2006). A recent model by Derdzinski et al. (2016) explored this possibility and investigated the location and mechanism by which dust grains can form and survive the intense UV and IR radiation in a nova explosion. They argued that the high energy particles accelerated at the shock have the potential to destroy the CO molecule that would allow for the formation of carbon-bearing dust grains in the carbon-poor, cool, dense shells following the shocked gas. Alternatively, dust condensation can proceed under non-equilibrium conditions such that only a small fraction of the carbon end up in the CO molecule (Evans et al. 1996), and lead to specific conditions where carbonaceous and oxide grains can condense simultaneously. Such non-equilibrium conditions can occur due to the presence of intermediate-mass elements, such as Al, Ca, Mg, or Si that may dramatically alter the stellar environment and allow the formation of carbon-rich dust even in a slightly O-rich environment (José et al. 2004, 2016). Finally, the timing of the formation of carbon- and oxygen-rich dust grains can be different. For example, carbon-rich dust was identified first, followed later by silicate formation in several novae (Gehrz et al. 1992; Evans et al. 1997; Mason et al. 1998; Sakon et al. 2016). These observations corroborate with the models of José et al. (2016) where chemical profiles with varying C/O ratios resulted in carbon-rich outer layers and oxygen-rich inner layers. Therefore, SiC grain condensation can occur via several ways namely, in the winds of carbon-rich outer layers, carbon-poor dust shells produced by particle irradiation, or proceed via kinetic effects during mixing of the different shells in the nova winds. The efficiency with which grain formation could occur in these situations are poorly constrained.

***4.3 SiC AB grains and C2.*** There are implications of this study for other types of SiC grains that have been classified in earlier work as SiC AB and C2 grains. SiC grains classified as Type C2s have carbon and nitrogen isotope compositions similar to the nova candidate grains ($^{12}C/^{13}C$ = 1–6.4; $^{14}N/^{15}N$ = 7–13) but have correlated enrichments in $^{29,30}Si$ compared to $^{28}Si$ (Liu et al. 2016). Our study included four grains as Type C2 (G278, G1342, GAB, G240-1), which were argued to not form in nova ejecta because Si isotopic signatures of these grains (Liu et al. 2016) did not agree with ONe nova models (José and Hernanz 2007; José et al. 2004). However, the new simulations described here fit these grains' carbon, nitrogen and sulfur isotopic compositions very well. However, three of these 4 grains (GAB, G1342, G249_1 in Figure 4) that have been measured for Al isotope ratios show low $^{26}Al/^{27}Al$ isotopic ratios that are not explained by the MDTNR models. The grain GAB, however, fits the MDTNR simulations (Table 3). Thus, a section of SiC C2 grains can have nova origins.

Another grain type whose origins need to be evaluated in the light of these new nova models are the SiC AB grains that have $^{15}N$ enrichments ($^{14}N/^{15}N$ ~ 4–200), low $^{12}C/^{13}C$ ratios ($^{12}C/^{13}C$ ~ 1–10), and enrichments in heavy Si isotopes up to about 200 ‰ (Amari et al. 2001). These grains typically show a very large range in $^{26}Al/^{27}Al$ ratios from ~$4\times10^{-5}$ to $2\times10^{-2}$. The carbon and nitrogen isotopic compositions of five AB grains namely M1-A5-G1424, M3-G1134, M3-G1332, M3-G319, and M3-G489 (Figure 1; Liu et al. 2017) can be explained by the 0.6–0.8$M_\odot$ MDTNR 25-75 CO nova simulations. Because these same grains have close-to-solar silicon isotope ratios, the same low mass MDTNR simulations can explain their silicon isotope compositions with 100% nova ejecta material (Figure 2). But these grains have $^{26}Al/^{27}Al$ ratios even lower than the Type C2 grains described above (Figure 4), and therefore their aluminum isotope compositions cannot be explained by the CO nova models.

Based on these observations of aluminum isotope ratios, we argue that grains whose carbon, nitrogen, silicon, and sulfur isotopes fall within the simulated models and those that exhibit $^{26}$Al/$^{27}$Al ratios >6×10$^{-2}$ are the most likely nova grains. We argue, however, that the $^{26}$Al/$^{27}$Al ratios alone cannot be diagnostic of nova origins, and therefore, we cannot rule out a nova origin for grains with low $^{26}$Al/$^{27}$Al ratios, observed in the SiC AB and C2 grains. Novae with an abundance of seed NeNaMgAl nuclei could produce large abundances of $^{26}$Al. $^{26}$Al synthesis requires peak temperatures on the order of (2-3)×10$^{8}$K (Ward and Fowler 1980) and such temperatures are achievable in the CO models described here. But if the abundance of Mg seed nuclei in CO WD is low, the production of $^{26}$Al can be hindered. Whether there are mechanisms that allow for low $^{26}$Al/$^{27}$Al ratios in CO nova need to be further investigated.

***4.4 Origin of grain SiC070 with known carbon, nitrogen, silicon and $^{4}$He/$^{20}$Ne isotope ratios.*** Finally, only one presolar grain SiC070, with low $^{12}$C/$^{13}$C ratio, along with He and Ne isotopic compositions has been reported in the literature to date (Heck et al. 2007). Both theoretical and observational evidence suggests that novae may be an important source of the radioactive isotope $^{22}$Na, which is involved in the production of the $^{22}$Ne (Ne-E) measured in SiC grains. The $^{4}$He/$^{20}$Ne ratio of the grain SiC070 is 60; its low $^{12}$C/$^{13}$C ratio of 3.5 was used to identify it as a SiC AB grain (Heck et al. 2007). Because the nova models described here, provide the He and Ne abundances, we compared the grain's carbon and noble gas compositions to the simulations. The composition of this grain can be explained best by the 1.15$M_\odot$ MDTNR 25-75 model: Mixing 96% of nova ejecta to 4% solar system material produces a $^{12}$C/$^{13}$C ratio =3.5 and assuming no solar system contribution for the He/Ne isotope ratios, we get $^{4}$He/$^{20}$Ne =67, which is very close to the grains' composition. However, neither its nitrogen or silicon isotope compositions can be explained by the 1.15$M_\odot$ MDTNR 25-75 or any other simulation. This gives us great confidence that this grain SiC070 is not of CO nova origin. Additional Ne isotopic measurements of gas rich SiC grains with low $^{12}$C/$^{13}$C ratios are required for suitable comparisons to the modeling in the future.

## 5. CONCLUSIONS

This work describes the results from new nova models in order to understand the progeny of presolar grains with low $^{12}C/^{13}C$ ratios (<10), low $^{14}N/^{15}N$ isotope ratios in the range ~5–20, large $^{30}Si$ excesses and high $^{26}Al/^{27}Al$ ratios. We explored several CO simulations where the mixing conditions were varied, and quantitative mixing models were carried out.

The simulations elucidate the puzzling isotopic compositions of the carbon-rich presolar grains. The comparisons of the simulations to the SiC grains show 2 major outcomes: First, the 0.8–1.35$M_\odot$ CO MDTNR 25-75 models, where mixing between the core and accreted material in the binary star system occurs after TNR is initiated, can explain the isotope compositions of four presolar grains quantitatively. One grain requires MDTNR 50-50 simulation with 1.35$M_\odot$ WD. Second, the grain compositions require <25% of solar system material to reproduce the grains' compositions, which confirm that nova dust grains are a component of the presolar grain inventory. Now that isotopic compositions of carbon, nitrogen, silicon, sulfur, and aluminum compositions in the C-rich nova grains have been constrained, further work is needed to understand the nature of the nova explosion and consequences of episodic mass loss. In addition, the stellar sites of most of the grains that don't match the CO nova simulations need to be investigated. For example, recent supernova models by Pignatari et al. (2015) that simulates core collapse supernova (CCSN) in a 25$M_\odot$, Z=0.02 progenitor star, where the star undergoes a rapid explosion at high kinetic energies of 4–7×$10^{51}$ ergs could be invoked to solve the origins of the SiC grains that are not nova condensates. This work is extremely relevant to the planetary science community because it enhances our understanding of the solar system environment prior to its formation. Injection of compositionally diverse presolar material from CO nova most probably occurred in the early solar system.

**Acknowledgements.** We would like to thank J. José and C. Iliadis for valuable discussions. SS also acknowledges partial support from NASA and HST grants to Arizona State University. This work was partially supported by startup funds to MB from Arizona State University. We would like to thank an anonymous reviewer for detailed comments that helped improve the manuscript immensely.

**References:**

Amari, S., Lewis, R. S., and Ander, E., 1994, GCA, 58, 459

Amari, S., & Zinner, E. 1997, in AIP Conf. Proc. 402, ed. T. J. Bernatowicz & E. Zinner (Woodbury, NY: AIP), 287

Amari, S., Gao, X., Nittler, L. R., et al. 2001, ApJ, 551, 1065

Anders, E., & Zinner, E., 1993, Meteoritics, 28, 490

Arnett, D., Meakin, C., and Young, P. A. 2010, ApJ, 710, 1619

Bernatowicz, T., Fraundorf, G., Tang, M., et al. 1987, Nature, 330, 728

Bode, M. F., and Evans, E., 2008, Classical Novae, Cambridge Univ. Press, 2008

Derdzinski, A. M., Metzger, B. D., & Lazzati, D., 2016, MNRAS, 469, 1314

Evans, A. et al., 1996, MNRAS, 282, 1049

Evans, A., Geballe T. R., Rawlings J. M. C. et al., 1996, MNRAS, 282, 192

Gao, X., & Nittler, L. R. 1997, Lunar Planet. Sci., 28, 393

Gehrz, R. D., Grasdalen, G. L., Greenhouse, M., et al. 1986 ApJ, 308, L63

Gehrz, R. D., Jones, T. J., Woodward, C. E., et al., 1992, ApJ, 400, 671

Gehrz, R. D., Greenhouse, M. A., Hayward, T. L., et al., 1995, ApJ, 448, L119

Heck, P., Marhas K. K., Hoppe, P., et al. 2007, 656, 1208

Hoppe, P., Strebel, R., Eberhardt, P., et al. 1996, Science, 272, 1314

Hoppe, P., & Ott U., 1997, in AIP Conf. Proc. 402, ed. T. J. Bernatowicz and E. Zinner (Woodbury, NY: AIP), 287

Hoppe, P., Leitner, J., Groner E. et al., 2010, ApJ, 719, 1370

Hoppe, P., Pignatari, M., Kodolanyi, J., et al. 2018, GCA, 221, 182

Iliadis, C., Downen, L. N., José, J., et al. 2018, ApJ, 855, 76

José, J., & Hernanz, M. 1998, ApJ, 494, 680

José, J., Coc, A., and Hernanz, M., 1999 ApJ, 520, 347

José, J., Hernanz, M., Amari, S., Lodders, K., & Zinner, E. 2004, ApJ, 612, 414

José, J., Halabi, G. M., F. El Eid, M., 2016, A&A, 593, A54

Liu, N., Nittler, L. R., Alexander, C. M. O'D., et al. 2016, ApJ, 820, 140

Liu, N., Nittler, L. R., Pignatari, M., et al. 2017, ApJ, 842, L1

Lodders, K., 2003, ApJ, 591, 1220

Lodders, K. & Palme, H. 2009, MAPS Suppl., 72, 5154

Mason, C. G., Gehrz, R. D., Woodward, C. E., et al. 1996 ApJ, 470, 577

Mason, C. G., Gehrz, R. D., Woodward, C. E., et al. 1997 ApJ, 470, 577

Mason, C. G., Gehrz, R. D., Woodward, C. E., et al. 1998 ApJ, 494, 783

Nittler, L. R., & Hoppe, P. 2005, ApJ, 631, L89

Nittler, L. R., & Cielsa, F. 2016, ARA&A, 54, 53

Nittler, L. R., Amari, S., Zinner, E., Woosley, S. E., and Lewis, R. S. 1996, ApJ, 462, L31

Nomoto, K., Thielemann, F. -K., & Wheeler J. C., 1984, ApJ, 279, L33

Parikh, A., Wimmer, K., Faestermann, T. et al 2014, Physics Letters B, 737, 314

Pignatari, M., Zinner, E., Hoppe, E., et al. 2015, ApJ, 808, L43

Rauscher, T., Heger, A., Hoffman, R. D., et al., 2002, ApJ, 576, 323

Sakon, I., Sako, S., Onaka T., et al., 2016, ApJ, 817, 145

Sallaska, A. L., Iliadis, C., Champagne, A. E., et al. 2013, ApJS, 207, 18

Starrfield, S., Gehrz, R. D. & Truran, J. W., 1997, in AIP Conf. Proc. 402, ed. T. J. Bernatowicz & E. Zinner (Woodbury, NY: AIP), 287

Starrfield, S., Iliadis, C., Hix, et al., 2009, ApJ, 692, 1532

Starrfield, S., Iliadis, C., & Hix, W. R., 2016, PASP 128, 051001

Starrfield, S., Timmes, F. X., Iliadis, C., et al. 2012, Baltic Astronomy, 21, 76

Starrfield S, Bose, M., Iliadis C., et al. 2018, ApJ in prep.

Whelan, J., & Icko, I., 1973, ApJ 186, 1007

Woosley, S. E., and Weaver, T. A. 1994, ApJ, 423, 371

Zinner, E. 2014, in Treatise on Geochemistry, Vol. 1, ed. A. M. Davis (2nd ed.; Oxford: Elsevier), 181

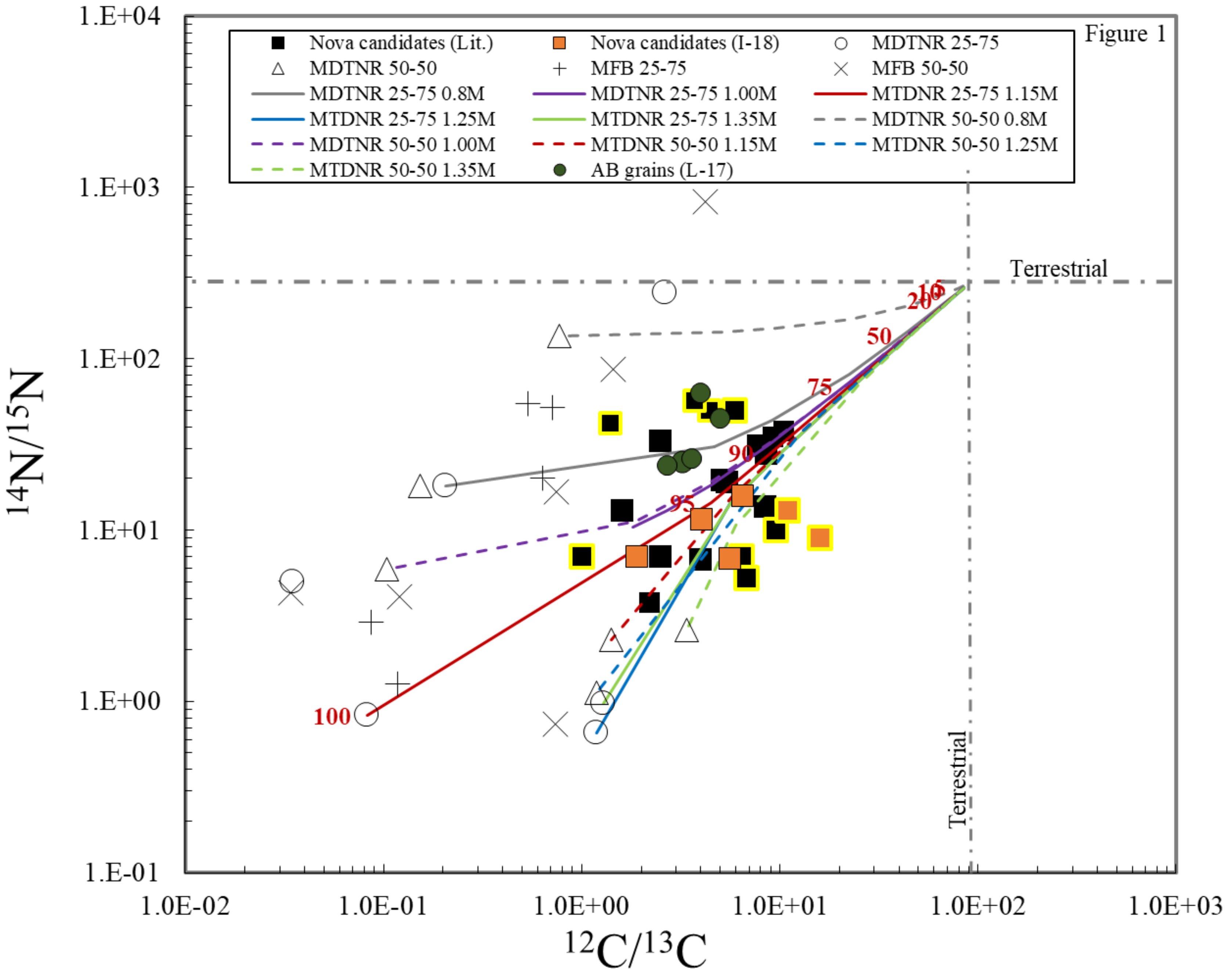

Figure 1: The $^{12}C/^{13}C$ and $^{14}N/^{15}N$ ratios of nova candidate grains from the literature (Hoppe et al. 1996, **2018**; Gao and Nittler 1997; Amari et al. 2001; Nittler and Alexander 2003; Nittler and Hoppe 2005; Liu et al. 2016, 2017) have been plotted in black. The SiC grains with a higher probability of being products of CO nova based on Iliadis et al. (2018) are shown in orange (I-18). This includes five SiC AB grains from Liu et al. (2017) and referred to as L-17. Black and orange grains with yellow outlines don't fit the simulations. Simulations from MDTNR and MFB models have also been plotted, and mixing lines between WD masses 0.8–1.35$M_{\odot}$ and solar system material are shown by lines of different colors. For the MDTNR model with WD mass of 1.15$M_{\odot}$, the proportion of nova ejecta is written next to the mixing lines. Except AF15bC-126-3, G240-1, G1697, Ag2, Ag2_6, G270_2, M2-A4-G672, M2-A5-G1211, M1-A8-G145, KJD-1-11-5 and KJD-3-23-5 (Table 2), all the grains can be quantitatively explained by the MDTNR 25-75 and 50-50 models.

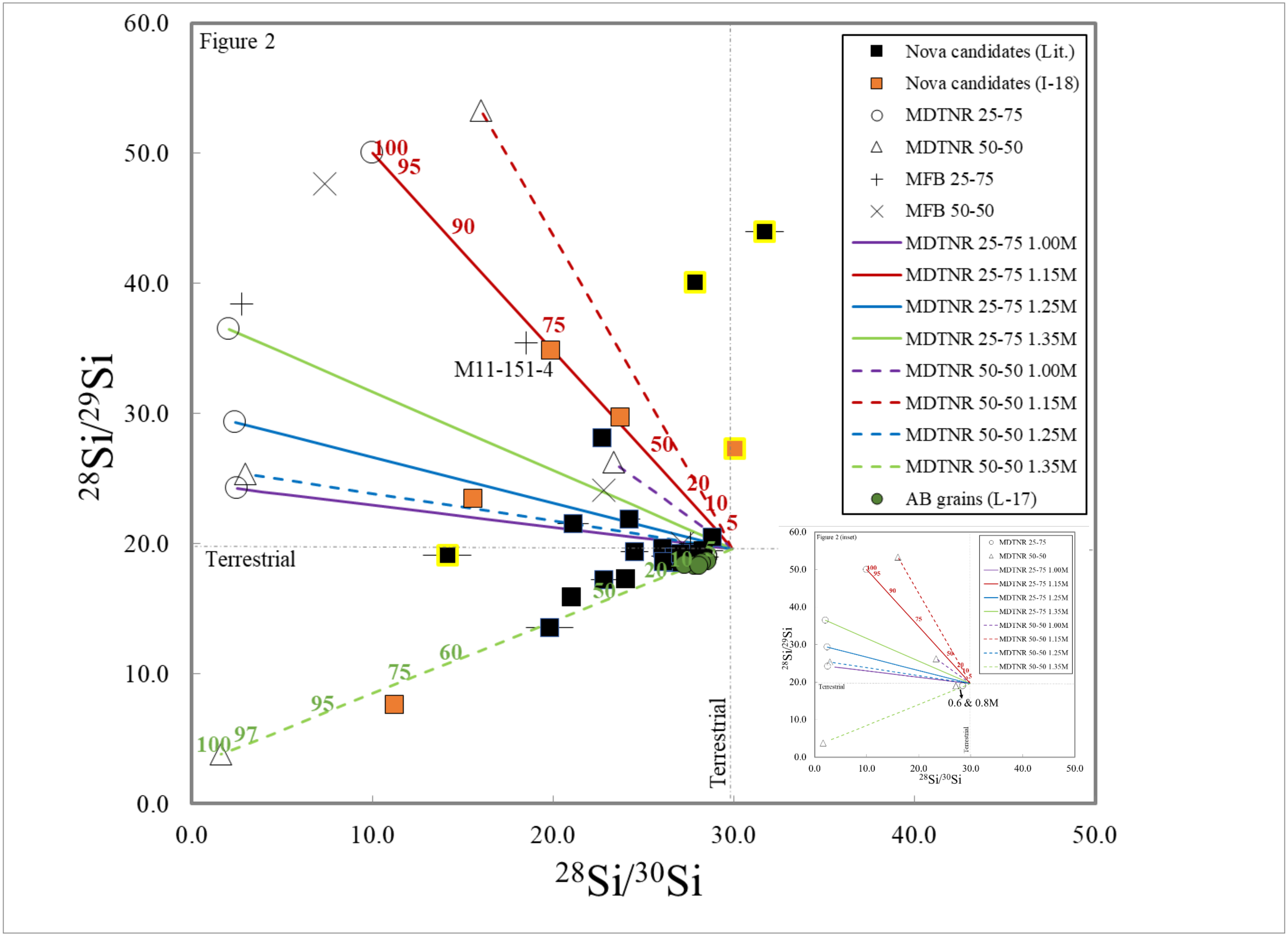

Figure 2: The $^{28}Si/^{29}Si$ and $^{28}Si/^{30}Si$ ratios of nova candidate grains from the literature (Hoppe et al. 1996, **2018**; Gao and Nittler 1997; Amari et al. 2001; Nittler and Alexander 2003; Nittler and Hoppe 2005; Liu et al. 2016, 2017) have been plotted in black. These ratios have been calculated by assuming terrestrial $^{29}Si/^{28}Si$ = 5.1% and $^{30}Si/^{28}Si$ = 3.34%. The SiC grains with a higher probability of being products of CO nova based on Iliadis et al. (2018) are shown in orange (I-18). This includes five SiC AB grains from Liu et al. (2017) and referred to as L-17. Black and orange grains with yellow outlines don't fit the simulations. Simulations from MDTNR and MFB models have also been plotted, and mixing lines between WD masses 1.15-1.35$M_{\odot}$ and solar system material are shown by lines of different colors. For the MDTNR model with 1.15$M_{\odot}$, the proportion of nova ejecta is written next to the mixing lines. Except G270_2, M2-A4-G27, AF15bB-429-3, M2-A5-G1211, all other grains can be well-described by the MDTNR 25-75 and 50-50 models. The inset shows the simulated compositions that are otherwise hidden by the grain data for clarity.

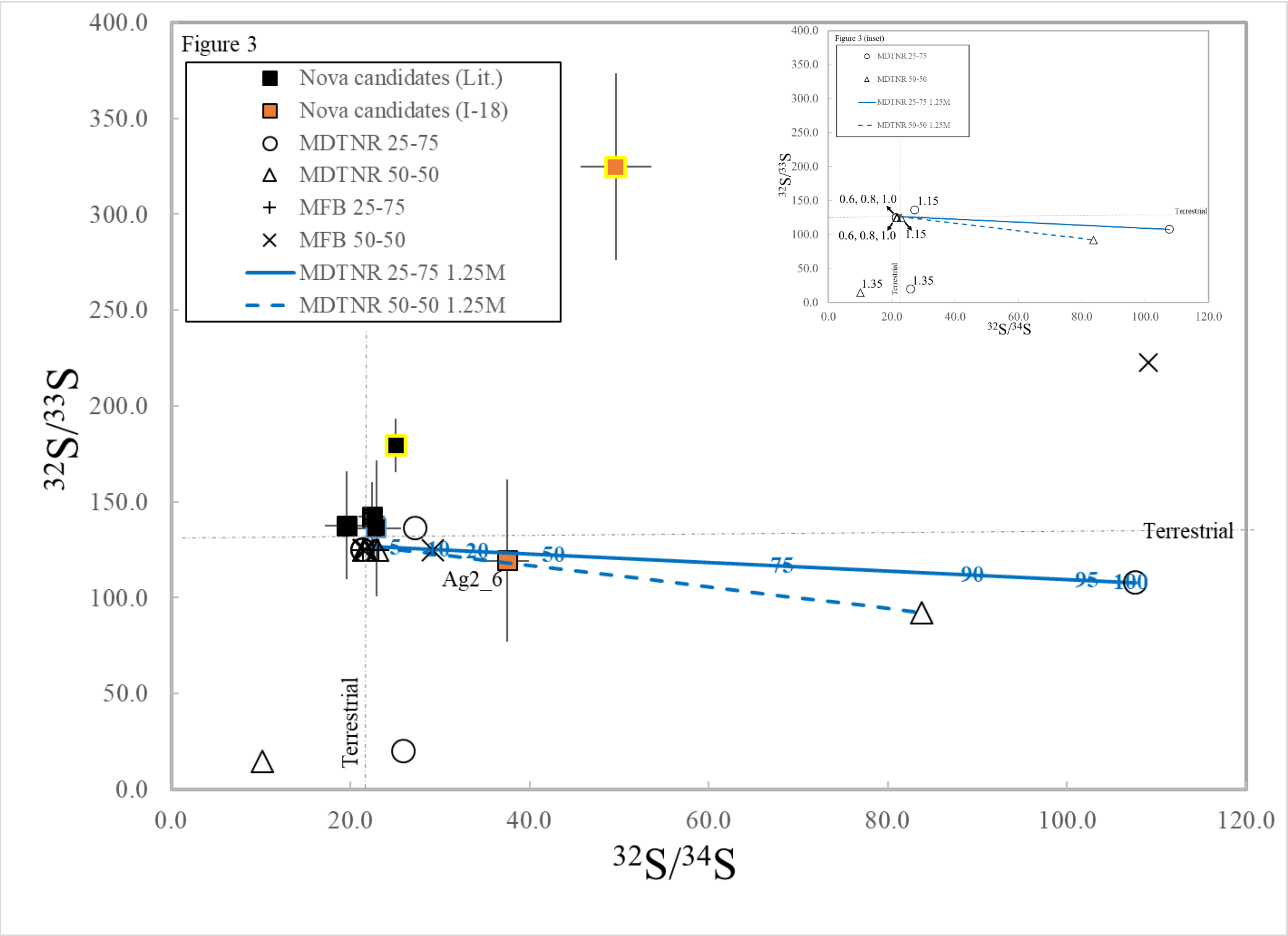

Figure 3: The $^{32}S/^{33}S$ and $^{32}Si/^{34}Si$ ratios of nova candidate grains from the literature (Hoppe et al. 1996; Gao and Nittler 1997; Amari et al. 2001; Nittler and Alexander 2003; Nittler and Hoppe 2005; Liu et al. 2016, 2017) have been plotted in black. These ratios have been calculated by terrestrial $^{33}S/^{32}S$ = 0.79 % and $^{34}S/^{32}S$ = 4.4 %. The SiC grains with a higher probability of being products of CO nova based on Iliadis et al. (2018) are shown in orange (I-18). Simulations from MDTNR models have also been plotted, and mixing lines between WD masses 1.25$M_\odot$ and solar system material are shown by blue lines. For the MDTNR model with 1.25$M_\odot$, the proportion of nova ejecta is written next to the mixing lines. One grain M1-A8-G145 with very large $^{33}S$ depletion is not plotted ($^{32}S/^{34}S$~40). Except for grains M1-A8-G145 and G270_2, remaining grains can be explained by MDTNR models. The inset shows the simulated compositions that are otherwise hidden by the grain data for clarity.

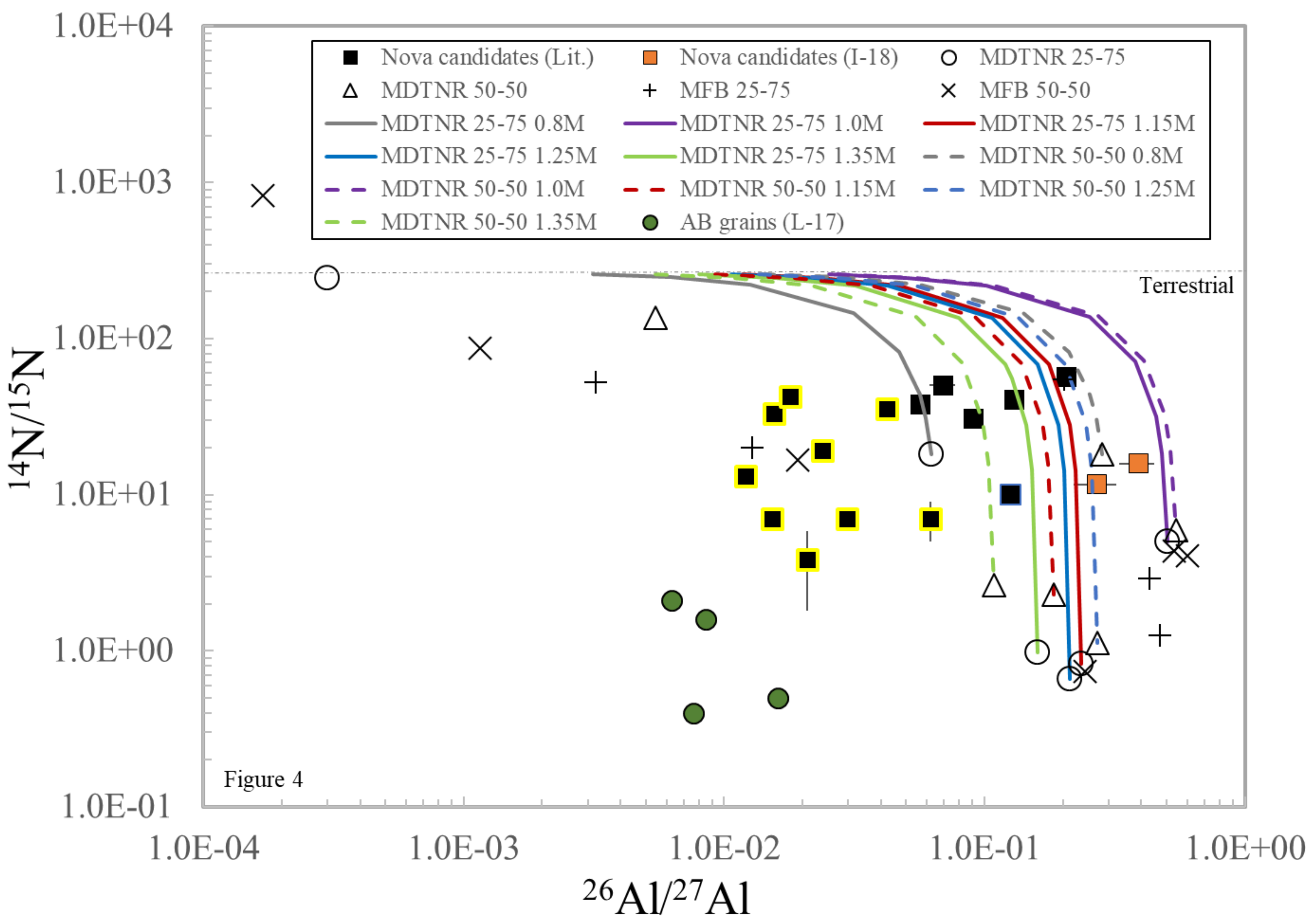

Figure 4

Figure 4: The nitrogen isotopic composition and $^{26}Al/^{27}Al$ ratios of nova candidate grains from the literature (Hoppe et al. 1996, **2018**; Gao and Nittler 1997; Amari et al. 2001; Nittler and Alexander 2003; Nittler and Hoppe 2005; Liu et al. 2016, 2017) have been plotted in black. The SiC grains with a higher probability of being products of CO nova based on Iliadis et al. (2018) are shown in orange (I-18). Remaining nova candidate grains and four SiC AB grains from Liu et al. (2017) and referred to as L-17 have much lower $^{26}Al/^{27}Al$ ratios. Simulations from MDTNR and MFB models have also been plotted, and mixing lines between WD masses 1.15–1.35$M_\odot$ and solar system material are shown by lines of different colors. Eight out of 17 grains can be quantitatively produced by the nova models described here.

**Table 1: Nova candidate SiC grains in the order of their discovery**

| **Grain** | **$^{12}C/^{13}C$** | **$^{14}N/^{15}N$** | **$\delta^{29}Si/^{28}Si$** | **$\delta^{30}Si/^{28}Si$** | **$\delta^{33}S/^{32}S$** | **$\delta^{34}S/^{32}S$** | **$^{26}Al/^{27}Al$** |
|---|---|---|---|---|---|---|---|
| KJC11[a] | 4.0±0.2 | 6.7±0.3 | … | … | … | … | … |
| KJGM4C-100-3[b] | 5.1±0.1 | 19.7±0.3 | 55±5 | 119±6 | … | … | 0.0114 |
| KJGM4C-311-6[b] | 8.4±0.1 | 13.7±0.1 | -4±5 | 149±6 | … | … | >0.08 |
| AF15bC-126-3[b] | 6.8±0.2 | 5.22±0.11 | -105±17 | 237±20 | … | … | … |
| AF15bB-429-3[b] | 9.4±0.2 | … | 28±30 | 1118±44 | … | … | … |
| M26a-53-8e[c] | 4.75±0.23 | … | 10±13 | 222±25 | … | … | … |
| **M11-151-4[d]** | **4.02±0.07** | **11.6±0.1** | **-438±9** | **510±18** | **…** | **…** | **0.27±0.05** |
| **M11-334-2[d]** | **6.48±0.08** | **15.8±0.2** | **-489±9** | **-491±18** | **…** | **…** | **0.39±0.06** |
| **M11-347-4[d]** | **5.59±0.13** | **6.8±0.2** | **-166±12** | **927±30** | **…** | **…** | **…** |
| G1342[e] | 6.4±0.08 | 7.0±0.14 | 445±34 | 513±43 | … | … | 0.015±0.001 |
| GAB[e] | 1.6±0.01 | 13.0±0.2 | 230±6 | 426±7 | -82±279 | -6±122 | 0.012±0.001 |
| G240-1[e] | 1.0±0.01 | 7.0±0.1 | 138±14 | 313±23 | … | … | 0.03±0.001 |
| G1748[e] | 5.4±0.02 | 19.0±0.2 | 21±4 | 83±5 | … | … | 0.024±0.004 |
| G283[e] | 12.0±0.1 | 41.0±0.5 | -15±3 | 75±4 | … | … | 0.13±0.04 |
| G1614[e] | 9.2±0.07 | 35±0.7 | 34±5 | 121±6 | … | … | 0.043±0.008 |
| G1697[e] | 2.5±0.01 | 33±0.8 | -42±12 | 40±15 | … | … | 0.016±0.004 |
| **G278[e]** | **1.9±0.03** | **7.0±0.2** | **1570±112** | **1673±138** | **…** | **…** | **…** |
| Ag2[e] | 2.5±0.1 | 7.0±0.1 | -304±26 | 319±38 | -92±222 | 162±106 | 0.062±0.01 |
| **Ag2_6[e]** | **16.0±0.4** | **9.0±0.1** | **-340±57** | **263±82** | **48±334** | **-394±106** | **…** |
| **G270_2[e]** | **11.0±0.3** | **13.0±0.3** | **-282±101** | **-3±131** | **-615±385** | **-542±175** | **…** |
| M2-A1-G410[f] | 10.4±0.3 | 38±0.5 | 19±13 | 88±17 | … | … | 0.057±0.001 |
| M2-A3-G581[f] | 7.8±0.22 | 31±2.4 | 52±11 | 147±12 | … | … | 0.09±0.003 |
| M2-A1-G114[f] | 16.2±0.5 | 56.0±2.0 | 20±25 | 107±35 | … | … | … |
| M2-A5-G269[f] | 8.5±0.19 | 28±1.3 | 7±10 | 59±13 | … | … | … |
| M2-A4-G672[f] | 9.6±0.3 | 10.0±0.8 | -90±18 | 419±28 | … | … | $(1.26\pm0.04)\times10^{-1}$ |
| M2-A4-G27[f] | 2.2±0.1 | 3.8±0.2 | -511±7 | 76±14 | … | … | $(2.09\pm0.79)\times10^{-2}$ |
| M2-A5-G1211[f] | 5.9±0.1 | 50.0±2.5 | -544±11 | -56±35 | … | … | … |
| M1-A8-G145[f] | 4.4±0.01 | 50.0±2.0 | 31±17 | 157±15 | -833±167 | -435±131 | $(6.93\pm0.09)\times10^{-2}$ |
| KJD-1-11-5[g] | 3.7±0.00 | 57.0±1.0 | -23±9 | 136±11 | -303±110 | -94±54 | $(0.21\pm1.2)\times10^{-2}$ |
| KJD-3-23-5[g] | 1.4±0.00 | 42.0±1.0 | 132±15 | 248±20 | -121±141 | 15±65 | $(1.8\pm0.2)\times10^{-2}$ |

a: Hoppe et al. (1996); b: Amari et al. (2001); c: Nittler and Alexander (2003); d: Nittler and Hoppe (2005); e: Liu et al. (2016); f: Liu et al. (2017); g: Hoppe et al. (2018)

High plausibility nova grains from Iliadis et al. (2018) are in bold.

**Table 2: Fits to the C, N, Si, S and Al isotopic data using new nova models**

| | MDTNR 25-75 | | | | | | MDTNR 50-50 | | | | | | No good fit(s) | No data |
|---|---|---|---|---|---|---|---|---|---|---|---|---|---|---|
| **Grain** | **1.35** | **1.25** | **1.15** | **1.00** | **0.8** | **0.6** | **1.35** | **1.25** | **1.15** | **1.00** | **0.8** | **0.6** | | |
| KJC11[a] | C,N | C,N | | | | | | C,N | | | | | | Si,S,Al |
| KJGM4C-100-3[b] | | | C,N | C,N | | | Si | | | C,N | Si | Si | | S,Al |
| KJGM4C-311-6[b] | | | | Si | | | C,N | Si | | | | | | S,Al |
| AF15bC-126-3[b] | | Si | | | | | | | | | | | C,N | S,Al |
| AF15bB-429-3[b] | | | | | | | | | | | | | Si | N,S,Al |
| M26a-53-8e[c] | | | | Si | | | | Si | | | | | | N,S,Al |
| **M11-151-4[d]** | | Al | **C,N,Si,Al** | | | | | Al | C,N | | | | | S |
| M11-334-2[d] | C,N | C,N | C,N | | | | | C,N | C,N | Al | Al | | Si | S |
| M11-347-4[d] | | | | | | | C,N | Si | | | | | | S,Al |
| *G1342[e]* | | | | | | | *C,N,Si* | | | | | | Al | S |
| *GAB[e]* | | | | *C,N,S* | S | S | Si | | S | *C,N,S* | | | Al | |
| G240-1[e] | | | | | | | Si | | | | | | C,N,Al | S |
| *G1748[e]* | | | C,N | *C,N,Si* | Si | Si | | Si | | C,N | Si | Si | Al | S |
| ***G283[e]*** | **C,N,Si,Al** | *C,N,Si* | C,N | *C,N,Si* | Si | Si | | *C,N,Si* | C,N | C,N | | | | S |
| G1614[e] | | | C,N | C,N | | | | | | C,N | Si | Si | Al | S |
| G1697[e] | Si | Si | | Si | | | | Si | | Si | | | C,N,Al | S |
| G278[e] | | | C,N | | | | Si | | | | | | | S,Al |
| Ag2[e] | | | Si | S | S | S | | | S | S | | | C,N,Al | |
| Ag2_6[e] | | S | Si | | | | | S | | | | | C,N | Al |
| G270_2[e] | | | | | | | | | | | | | C,N,Si | S,Al |
| ***M2-A1-G410[f]*** | | | C,N | *C,N,Si* | **C,N,Si,Al** | Si | | | C,N | C,N | Si | Si | | S |
| M2-A3-G581[f] | | | C,N | C,N | | | | Al | | C,N | Si | Si | | S |
| *M2-A1-G114[f]* | C,N | CN | C,N | *C,N,Si* | C,N | | | C,N | C,N | C,N | Si | Si | | S,Al |
| *M2-A5-G269[f]* | C,N | *C,N,Si* | C,N | *C,N,Si* | Si | Si | | *C,N,Si* | C,N | C,N | | | | S,Al |
| M2-A4-G672[f] | Al | | | Si | | | | Si | | | | | C,N | S |
| M2-A4-G27[f] | | | | | | | | C,N | C,N | | | | Si,Al | S |
| M2-A5-G1211[f] | | | | | | | | | | | | | C,N,Si | S,Al |
| M1-A8-G145[f] | | | | | Al | | | | | | Si | Si | C,N,S | |
| KJD-1-11-5[g] | | | | | | | | | | | Si | Si | C,N,Al | |
| KJD-3-23-5[g] | | | | | | | Si | | | | | | C,N,Al | |

Grains where 5 isotopic ratios are a good match are in **bold**. Grains where 4 isotopic ratios are a good match are in *italics*. Si and S isotope ratios comprise of two ratios each.

a: Hoppe et al. (1996); b: Amari et al. (2001); c: Nittler and Alexander (2003); d: Nittler and Hoppe (2005); e: Liu et al. (2016); f: Liu et al. (2017); g: Hoppe et al. (2018)

**Table 3: Proportion of Nova matter for a Subset of Nova candidate grains**

| | Best fits for C, N, Si and S | % of nova matter for C/N | % of nova matter for Si | % of nova matter for S | % of nova matter for Al | Nova grain |
|---|---|---|---|---|---|---|
| **M11-151-4[d]** | **MDTNR 25-75 1.15M** | **95** | **75** | **No S data** | **96** | **yes** |
| **M2-A1-G410[f]** | MDTNR 25-75 1.00M | 88 | 20 | No S data | No good fit for Al | no |
| | **MDTNR 25-75 0.8M** | **90** | **100** | | **92** | **yes** |
| **G1342[e]** | **MDTNR 50-50 1.35M** | **98** | **55** | **No S data** | **No good fit for Al** | **maybe?** |
| **GAB[e]** | **MDTNR 25-75 1.00M** | **98** | **No Si fit** | **100** | **No good fit for Al** | **yes** |
| | **MDTNR 50-50 1.00M** | **98** | **No Si fit** | **100** | **No good fit for Al** | **yes** |
| **G283[e]** | **MDTNR 25-75 1.35M** | **87** | **20** | **No S data** | **85** | **yes** |
| | MDTNR 25-75 1.25M | 87 | 20 | | No good fit for Al | no |
| | MDTNR 50-50 1.25M | 87 | 20 | | No good fit for Al | no |
| | MDTNR 25-75 1.00M | 85 | 10 | | No good fit for Al | no |
| G1748[e] | MDTNR 25-75 1.00M | 95 | 20 | No S data | No good fit for Al | no |
| M2-A1-G114[f] | MDTNR 25-75 1.00M | 80 | 20 | No S data | No Al data | no |
| M2-A5-G269[f] | MDTNR 25-75 1.00M | 90 | 10 | No S data | No Al data | no |
| | MDTNR 25-75 1.25M | 90 | 10 | | | no |
| | MDTNR 50-50 1.25M | 90 | 10 | | | no |

Grains in bold are nova grains

d: Nittler and Hoppe (2005); e: Liu et al. (2016); f: Liu et al. (2017)